\documentstyle[12pt,aas2pp4]{article}

\begin{document}

\title{A Search for Hard X-Ray Emission from Globular Clusters - 
Constraints from BATSE}
\author{E. Ford\altaffilmark{1}, P. Kaaret\altaffilmark{1},
        B.A. Harmon\altaffilmark{2}, M. Tavani\altaffilmark{1},
        and S.N. Zhang\altaffilmark{3}}
\altaffiltext{1}{Columbia University, Department of Physics and
 Columbia Astrophysics Lab, 538 W. 120th Street, New York, NY 10027}
\altaffiltext{2}{NASA/Marshall Space Flight Center, 
 ES 66, Huntsville, AL 35812}
\altaffiltext{3}{University Space Research Association/ MSFC,
 ES 66, Huntsville, AL 35812}

\begin{abstract}
We have monitored a sample of 27 nearby globular clusters in the hard X-ray 
band ($20-120$~keV) for $\sim$1400 days using the BATSE instrument on board 
the Compton Gamma-Ray Observatory.  Globular clusters may contain a large 
number of compact objects (e.g., pulsars or X-ray binaries containing 
neutron stars) which can produce hard X-ray emission. Our 
search provides a sensitive ($\sim$50 mCrab) monitor for hard X-ray transient 
events on time scales of $\gtrsim1$~day and a means for observing persistent 
hard X-ray emission. We have discovered no transient events from any of the 
clusters and no persistent emission. Our observations include a sensitive 
search of four nearby clusters containing dim X-ray sources: 47~Tucanae, 
NGC~5139, NGC~6397, and NGC~6752. The non-detection in these clusters implies a 
lower limit for the recurrence time of transients of 2 to 6~years for events 
with luminosities $\gtrsim10^{36}$~erg~s$^{-1}$ ($20-120$~keV) and 
$\sim20$~years if the sources in these clusters are taken collectively.
This suggests that the dim X-ray sources in these clusters are not transients 
similar to Aql~X-1. We also place upper limits on the persistent emission in 
the range $(2-10)\times10^{34}$~erg~s$^{-1}$ ($2\sigma$, $20-120$~keV) for 
these four clusters. For 47~Tuc the upper limit is more sensitive than previous
measurements by a factor of 3.  We find a model dependent upper limit of 19 
isolated millisecond pulsars (MSPs) producing gamma-rays in 47~Tuc,
compared to the 11 observed radio MSPs in this cluster.

\end{abstract}

\keywords{clusters:globular, 47~Tucanae, NGC~5139, NGC~6752, NGC~6397,
X-ray:transients}

\section{Introduction}

Globular clusters (GCs) are remarkable stellar systems where a variety of 
compact objects may form and evolve. Numerous millisecond pulsars (MSPs) 
inhabit GCs as revealed by radio observations (Lyne 1996), e.g. 11 MSPs in 
47~Tuc (Robinson et al. 1995). In the X-ray band, the number 
of bright sources ($1-10$~keV) per unit stellar mass is two orders of 
magnitude larger for GCs than for the remainder of the galaxy (Clark 1975). 
Persistent X-ray sources in GCs divide into two luminosity groups: one with
low luminosities ($L_{x}\lesssim 10^{34.5}$~erg~s$^{-1}$, $0.5-4.5$~keV), and 
the other with high luminosities ($L_{x}\gtrsim10^{36}$~erg~s$^{-1}$) 
(Hertz and Grindlay 1983, Hertz and Wood 1985, Verbunt et al. 1995). The 
nature of the low luminosity, dim X-ray sources (DXSs) has been controversial 
(for reviews see Bailyn 1995, Verbunt et al. 1994, Grindlay 1993b, 
Hut et al. 1992). Suggestions for their origin include cataclysmic variables, 
with mass accreting onto a white dwarf (Grindlay 1993a), low mass X-ray 
binaries (LMXBs) with accretion at quiescent levels onto a neutron star (NS) 
or black hole (BH) (Verbunt, Van Paradijs and Elson 1984), plerionic MSP 
binaries (Tavani 1991) or chance superpositions of foreground/background 
objects (Margon and Bolte 1987).

% previous transients

We have employed the BATSE instrument on board the Compton Gamma-Ray 
Observatory to search for transient and persistent hard X-ray emission 
($\gtrsim20$~keV). Hard X-rays may be produced by GC compact objects in a 
number of ways. In a binary, hard X-ray emission can be powered by mass 
accretion onto the BH/NS primary, or in pulsars, from shock interactions in 
the relativistic pulsar winds in the binary. Hard X-ray emission may also be 
produced by magnetospheric emission of isolated MSPs. Transient emission has 
been observed previously from GCs. NGC~6440, a GC containing DXSs, has 
exhibited a high luminosity transient episode (Forman, Jones and Tananbaum 
1976). M15 may also have shown a transient event (Pye and McHardy 1983).
In the hard X-ray band, SIGMA has observed transient emission from the 
globular cluster Terzan~2. Recent ROSAT HRI observations indicate that the 
SIGMA source is most likely associated with the X-ray burster XB~1724-30
(Mereghetti et al. 1995). SLX~1732-304 in the GC Terzan~1 has also been 
detected as a hard X-ray source (Churazov et al. 1994). Our search does not 
include Terzan~2, Terzan~1 or NGC~6440 due to problems with nearby interfering
sources. We have been able to observe M15 however.

% our search and outline of the rest of the paper

In Section 2, we describe hard X-ray emission mechanisms in detail. 
Section 3 outlines the BATSE earth occultation flux measurement technique 
and the parameters of our search. In Sections 4 and 5 we present the results 
of our search, providing upper limits on transient events and constraints on 
the recurrence times of transients from DXSs. Section 6 contains upper limits 
on persistent emission and constraints on the number of isolated and 
interacting MSPs. Section 7 is a summary and conclusion.

\section{Hard X-ray emission from compact objects}

\subsection{Accreting black holes}

Soft X-ray transients (SXTs) have exhibited emission in excess of 10~keV 
(e.g. White, Kaluzienski and Swank 1984). These hard power law tails along 
with an anticorrelation of intensity and spectral hardness have been taken as 
indications of a BH primary (Tanaka 1989, Tanaka and Lewin 1995). 

Originally, the discovery of X-ray sources in GCs was interpreted as a possible 
manifestation of massive accreting BHs (e.g., Bahcall and Ostriker 1975).  
We know today that {\it all} bright X-ray sources in GCs show X-ray bursts, 
and therefore most likely contain neutron stars. Though it is unlikely that 
clusters contain very massive black holes, they may harbor a population of 
stellar mass black holes. Kulkarni, Hut and McMillan (1993) estimate that 
there may be $\sim10$~systems in galactic GCs in which a BH of mass 
$\sim10$M$_\odot$ has captured a companion. In a few of these systems the 
companion may have had time to evolve off the main sequence, producing an 
active X-ray binary. Hard X-ray activity from such systems would be observable 
in our search.

\subsection{Accreting neutron stars}

LMXBs that contain NS primaries and accrete with luminosities below 
some critical value also exhibit transient episodes of hard X-ray emission 
(e.g. Barret and Vedrenne 1994) and an anticorrelation of spectral hardness 
and intensity, e.g.: 4U0614+091 (Barret and Grindlay 1995a), 
4U1608-522 (Mitsuda et al. 1989). 

\placetable{table:hardx}

\begin{table*}[tbp]
\begin{tabular}{|l|l|l|l|}
\tableline 
Source & D     & $L_x$   & Observation \\
       & (kpc) & (erg s$^{-1}$) &             \\ \tableline
Aql~X-1 & 2.5 & $1\times 10^{36}$ & BATSE ($20-120$ keV)\tablenotemark{a} \\ 
  \tableline
Cen~X-4 & 1.2 & $7\times 10^{36}$ & Signe 2MP ($13-163$ keV) \tablenotemark{b} \\ 
  \tableline
XB 1724-308 (Ter~2) & 14.0 & $2\times 10^{37}$ & SIGMA ($35-200$ keV) %
  \tablenotemark{c} \\
  \tableline
KS1731-260 & 8.5 & $2\times 10^{37}$ & SIGMA \tablenotemark{d} \\ 
  \tableline
4U1608-522 & 3.6 & $6\times 10^{36}$ & BATSE\tablenotemark{e} \\ 
  \tableline
A1742-294 & 8.5 & $6\times 10^{36}$ & SIGMA\tablenotemark{f} \\
  \tableline
\end{tabular}
\tablenotetext{a}{(Harmon et al. 1996)} %
\tablenotetext{b}{(Bouchacourt et al. 1984)} %
\tablenotetext{c}{(Barret et al. 1991)} %
\tablenotetext{d}{(Barret et al. 1992)} %
\tablenotetext{e}{(Zhang et al. 1996)} %
\tablenotetext{f}{(Churazov et al. 1995)} %
\caption{Hard X-ray outburst luminosities of five X-ray bursters. $L_x$ is the 
peak outburst luminosity for the $20-120$~keV band. The energy bands of each
instrument are shown.}
\label{table:hardx}
\end{table*}

Table~\ref{table:hardx} lists the observed hard X-ray properties of several 
transient and persistent LMXBs. All of these sources have been observed as 
type-I X-ray bursters which indicate NS primaries. The hard X-ray outbursts 
generally exceed $10^{36}$~erg~s$^{-1}$, with emission extending above 100~keV. 
The time scales for the hard X-ray outbursts are typically $10-100$~days. Such 
events would be easily detectable in our search. Additional bursters are being 
detected as hard X-ray sources in a BATSE monitoring program 
(Barret et al. 1996).

\subsection{Isolated and binary Pulsars}
% persistent hard X-rays and MSPs

Persistent hard X-ray emission may also be produced by MSPs, and the fact that
GCs contain large numbers of MSPs makes them ideal systems in which to search 
for such emission (Chen 1991, Tavani 1993b). Recent theoretical work has shown 
that hard X-ray and gamma ray emission may be produced by MSPs either in 
isolation (Chen, Middleditch and Ruderman 1993) or in interacting binaries 
(Tavani 1993a). Isolated MSPs could produce hard X-ray emission by the same 
mechanism operating in young Crab-like pulsars. Though the magnetic fields are 
smaller in recycled pulsars, the voltage drop at the light cylinder may be 
similar due to the smaller rotation periods (Ruderman and Cheng 1988). In 
binary systems, MSPs may emit hard X-rays by interaction with a companion 
(Tavani 1991). This is in analogy to PSR~B1957+20 in which the pulsar is 
evaporating its companion (e.g. Kluzniak et al. 1988). In such systems the 
pulsar is partly or fully buried in a gaseous envelope originating from its 
companion. The interaction of the MSP wind with the surrounding material 
produces energetic emission by synchrotron radiation or inverse Compton 
scattering of shock accelerated particles. Emission, which can be explained by 
this model, has recently been observed in the $1-200$~keV range from the 
binary pulsar PSR~1259-63 near periastron (Grove et al. 1995, 
Tavani et al. 1996).

\section{Cluster Search with BATSE}

% *** brief BATSE description

The BATSE instrument was designed to provide continuous coverage of the
entire sky in the hard X-ray/gamma-ray band from $20$~keV $-$ 2~MeV 
(Harmon et al. 1992), and has an optimal sensitivity in the range $20-120$~keV
for a source with a typical photon index of 2.
Source fluxes are measured by BATSE using the earth as an occulting object.
As a source rises or sets, steps appears in the detector count rates due to
attenuation by the Earth's atmosphere. Source fluxes are determined by fitting 
the amplitude of these steps. Thus, depending on the geometry of the 
Earth's rising and setting limbs, one can make two source measurements per GRO 
orbit (approximately 90 min.), and from these construct a light curve.

Fluxes in the resulting light curve, for a location in the sky with no sources,
form a distribution about a mean flux of zero. Both positive and negative 
fluxes are produced in the occultation edge fits, all with formal error 
estimates. The distribution of flux estimates has a width which is larger than 
expected for a Poisson distribution of background rates. This is due to 
systematic effects, such as contributions in the occultation fits from nearby 
sources and spurious background fluctuations.

The main limitation of the occultation technique is source confusion due to 
the large angular extent of the Earth's occulting limbs. There are several 
diagnostics that can be employed to identify confused sources. One such check 
is to compare the measured relative rates in separate detectors to the rates 
expected given the BATSE detector responses (Pendleton et al. 1996). Another 
technique is to compare the rates from the rising limb to those from the 
setting limb. For constant flux from the expected source, the rising and 
setting edges should give the same rates. The precession of the earth's limb 
can also be used to localize the emission, a technique that has been 
successfully employed to create images of sky regions for brighter, long 
duration events (Zhang et al. 1993).

% cluster list and motivation

\begin{figure*}
\figurenum{1a}
\epsscale{1.8}
\plotone{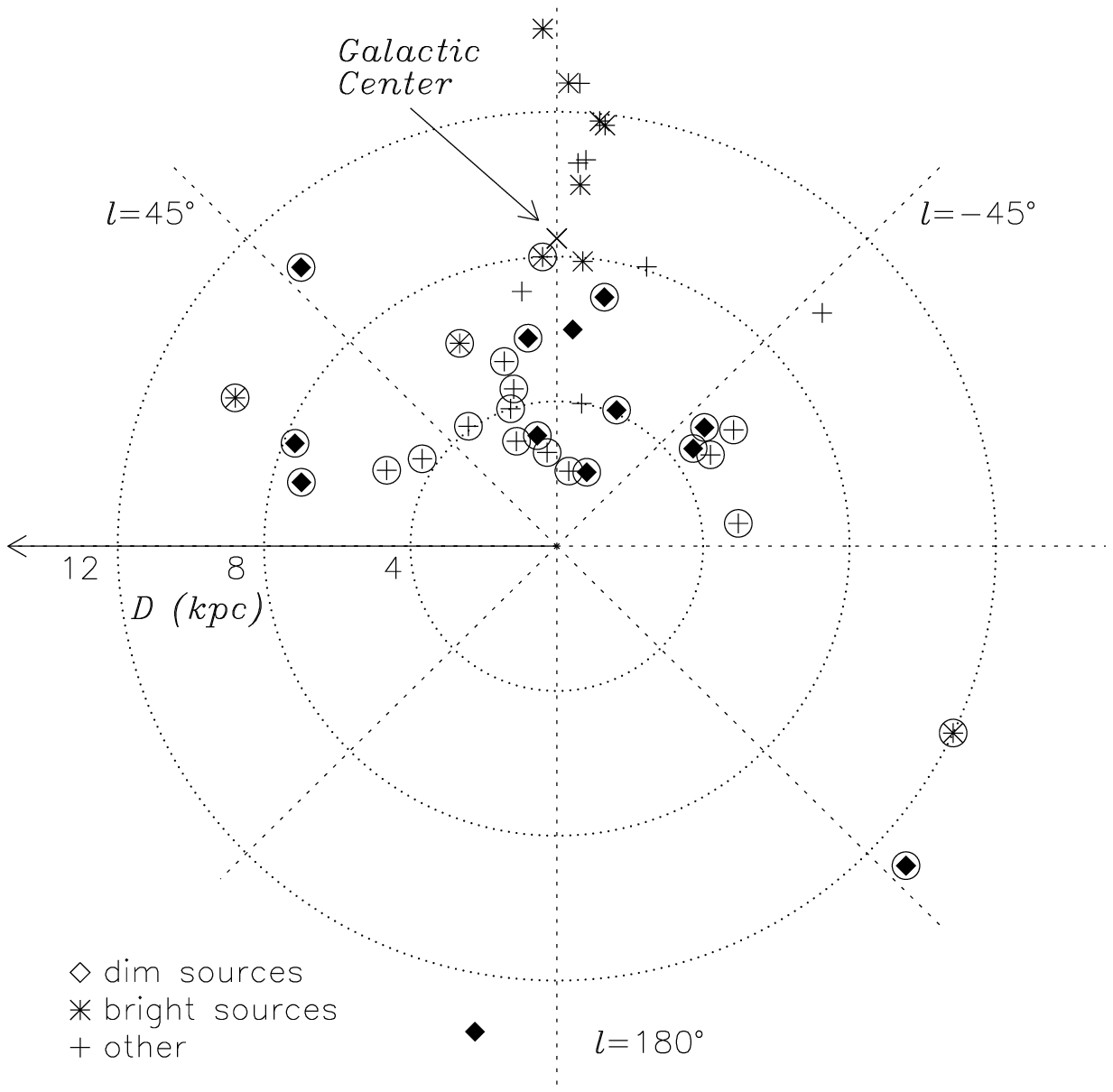}
\caption{Projection of the positions of globular clusters onto the galactic 
plane. All the circled clusters are included in our Search~A or Search~B.}
\label{fig:proj_full}
\end{figure*}

\begin{figure*}
\figurenum{1b}
\epsscale{2.1}
\plotone{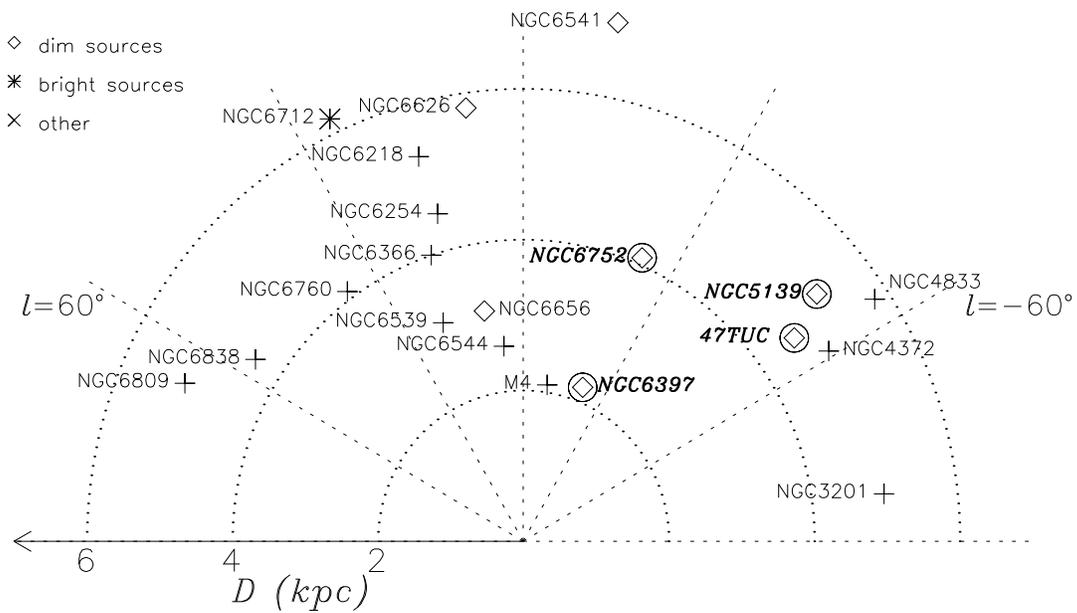}
\caption{Globular clusters within 6 kpc projected onto the plane. The four
clusters for which we have performed the sensitive Search~B are circled. All 
the other clusters are included in Search~A.}
\label{fig:proj}
\end{figure*}

\placefigure{fig:proj_full}
\placefigure{fig:proj}

We divide our search of 27 GCs into two parts, which we will refer to as
Search~A and Search~B (see Figure~\ref{fig:proj_full}). Search~A is a lower
sensitivity search of 23 GCs which are at larger distances and/or in 
regions of the sky prone to source confusion. Search~A includes GCs at 
distances $\lesssim6$~kpc, GCs near the galactic center, GCs containing no 
known DXSs, and GCs known to contain bright LMXBs. Search~B is a higher 
sensitivity search of four nearby GCs which are not subject to serious source 
confusion problems. We limit this search to GCs at $\lesssim6$~kpc which also 
contain DXSs near the cluster core. We also exclude clusters within 
15$^{\rm{o}}$ of the galactic center, to avoid source confusion in this dense 
area. The GCs of Search~B are shown in Figure~\ref{fig:proj} (circled). In 
what follows, all cluster parameters have been taken from Peterson (1993).

\section{Results from Search~A}

\begin{table*}[tbp]
\begin{tabular}{|l|l|l|l|l|l|l|}
\tableline
Cluster & $D$   & $R_{ul}$ & $F_{ul}$ & $L_{ul}$  & coverage & Persistent \\
 & (kpc) & (cnt s$^{-1}$) & ($\gamma$ cm$^{-2}$s$^{-1})$ & (erg s$^{-1}$) %
 & (\%) & X-ray \\
 &  &  & [$\times 10^{-2}$] & [$\times10^{36}$] &  & Source(s) \\ \tableline

M4      &  2.0 & 3.5 & 2.8 &  0.9 & 80.0 &  \\
NGC6544 &  2.5 & 4.0 & 3.2 &  1.7 & 77.1 &  \\
NGC6656 &  3.0 & 2.6 & 2.1 &  1.6 & 82.4 & dim \\
NGC6838 &  3.9 & 4.0 & 3.2 &  4.0 & 80.5 &  \\
NGC6539 &  4.0 & 4.4 & 3.5 &  4.7 & 70.9 &  \\
NGC6366 &  4.0 & 6.5 & 5.2 &  6.9 & 74.9 &  \\
NGC6760 &  4.2 & 6.3 & 5.0 &  7.2 & 38.7 &  \\
NGC6254 &  4.3 & 3.3 & 2.6 &  4.0 & 74.3 &  \\
NGC6809 &  4.8 & 4.5 & 3.6 &  6.8 & 72.3 &  \\
NGC3201 &  5.0 & 7.0 & 5.6 & 11.6 & 77.6 &  \\
NGC4372 &  5.2 & 4.0 & 3.2 &  7.2 & 69.8 &  \\
NGC6218 &  5.6 & 3.8 & 3.0 &  7.8 & 74.5 &  \\
NGC4833 &  5.8 & 2.9 & 2.3 &  6.4 & 69.0 &  \\
NGC6626 &  5.9 & 4.0 & 3.2 &  9.2 & 79.4 & dim \\
NGC6541 &  6.6 & 3.3 & 2.6 &  9.3 & 73.3 & dim \\
NGC6712 &  6.8 & 4.4 & 3.5 & 13.4 & 72.9 & bright \\
NGC7099 &  7.4 & 3.0 & 2.4 & 10.9 & 88.4 & dim \\
NGC6341 &  7.5 & 3.5 & 2.8 & 13.0 & 85.7 & dim \\
NGC6624 &  8.1 & 3.5 & 2.8 & 15.2 & 84.2 & bright \\
NGC5272 & 10.1 & 2.4 & 1.9 & 16.0 & 88.0 & dim \\ 
M15     & 10.5 & 3.5 & 2.8 & 25.5 & 87.6 & bright \\
NGC1851 & 12.2 & 4.5 & 3.6 & 44.3 & 88.1 & bright \\
NGC1904 & 13.0 & 3.1 & 2.5 & 34.9 & 88.1 & dim \\ \tableline
\end{tabular}
\caption{Event upper limits and coverage time for the 23 GCs in Search~A.
$R_{ul}$ is the raw count rate upper limit for hard X-ray 
transient events, and $F_{ul}$ the approximate corresponding limit in flux 
units. The luminosity upper limit ($20-120$~keV), $L_{ul}$, assumes the stated
distances and power law spectra with index 2. The total time span is 
1630 days for NGC~6544 and NGC~6656 and 1490 days for the other clusters. The 
clusters containing bright and dim X-ray sources are marked.}
\label{table:roughlimits}
\end{table*}

\placetable{table:roughlimits}

We have generated light curves for all the clusters extending from April~1991 
to March~1995 ($\sim$1400 days). For Search~A, we have analyzed the rate 
history of each cluster for events with structure on a time scale of one or 
more days. Generally in each light curve there are several ``events" apparent; 
most of these can be attributed to interference from nearby bright sources. 
There are, however, a few spikes that cannot readily be explained as 
interfering sources. None of these events are particularly outstanding in 
terms of amplitude or structure. For Search~B we have further analyzed such 
events to determine if they originate from the clusters, as described below. 
For the 23 clusters in Search~A, however, such an analysis is impractical. For 
Search~A, therefore, we take the largest of the outstanding features to 
set conservative upper limits to the minimum observable event amplitude 
(Table~\ref{table:roughlimits}). We initially find count rate upper limits
($R_{ul}$ in Table~\ref{table:roughlimits}). To convert these limits to photon 
fluxes, we have multiplied by a constant conversion factor determined from 
other light curves for which we have properly deconvolved the instrument 
response with an assumed photon index of 2.0 to generate photon flux histories. 
This conversion is accurate to within approximately $30\%$. From the photon 
rate, we calculate a luminosity upper limit by assuming power law spectra with 
index 2. The effective time coverage for each cluster is shown in 
Table~\ref{table:roughlimits}. Coverage with BATSE is less than $100\%$ and 
varies as a function of sky position due to the geometry of the Earth's 
occulting limbs and due to interfering sources. When target occultations occur 
within 10 seconds of the occultation of a bright interfering source, no edge 
fits are produced. This leads to data gaps, as in the case of NGC~6760, which 
has highly non-uniform coverage due to the nearby source Aql~X-1. Note that we 
obtain a limit of $\sim10^{37}$~erg~s$^{-1}$ for NGC~5272 (M3) which is known 
to contain a time variable supersoft X-ray source. Note also that the cluster
NGC~6440 has been excluded from our search due to nearby interfering sources.

\section{Results from Search~B}

For Search~B, we have produced rate histories over the time period mentioned 
above and properly deconvolved the detector response to obtain photon light 
curves. We have searched these light curves for outburst events. Candidate 
events are identified as features of duration one or more days which rise at 
least $1.5\sigma$ above the background flux distribution. With this criterion 
we have identified 40 possible candidate events. We analyzed each of these 
isolated events further to determine if they originate from the clusters.
We check for nearby interfering sources and compare the relative rates in 
the detectors. This analysis eliminates all but 13 of the events as candidates. 
These 13 candidates consist of peak-like structures with durations of 1 to 8
days. The significance of these events, as measured from the photon flux
error bars near the peaks, ranges from $2\sigma$ to $6\sigma$. The maximum 
luminosities in these candidate events would range from 
$0.7-6\times10^{36}$~erg~s$^{-1}$, and in one case 
$9\times10^{36}$~erg~s$^{-1}$ ($20-120$~keV). These luminosities are similar 
to the expected transient peak luminosities (c.f. Table~\ref{table:hardx}) and 
deserve further investigation.

\begin{figure*}
\figurenum{2a}
\epsscale{1.5}
\plotone{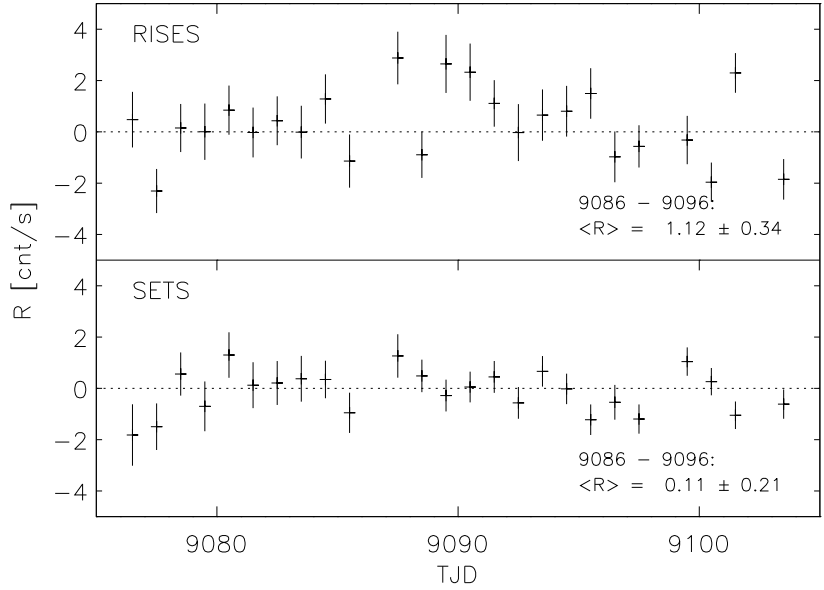}
\caption{Rate histories during candidate event in NGC~6752 near TJD~9090. 
The rate histories have been separated into rise and set rates, and corrected 
approximately for detector response. The average rise and set rates during
the approximate time of the event are shown.}
\label{fig:NGC6752cand}
\end{figure*}

\begin{figure*}
\figurenum{2b}
\epsscale{1.5}
\plotone{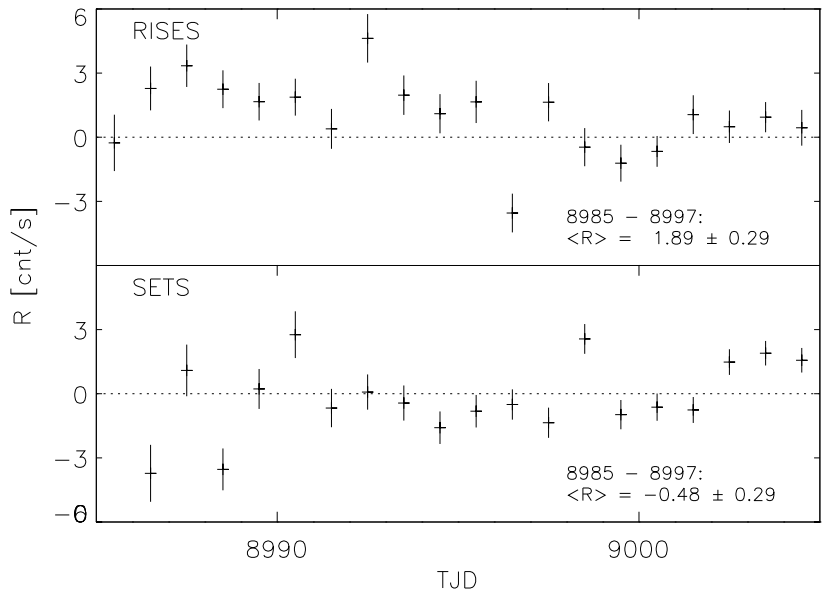}
\caption{Rate histories during candidate event in NGC~5139 near TJD~8990.}
\label{fig:NGC5139cand}
\end{figure*}

\placefigure{fig:NGC6752cand}
\placefigure{fig:NGC5139cand}

We further analyze candidate events by studying the correlation between 
the rising and setting rates within a given GRO pointing period. 
Figures~\ref{fig:NGC6752cand} and \ref{fig:NGC5139cand} show example
rate histories for candidate events in NGC~6752 and NGC~5139. For each source, 
the rise and set rates clearly have different temporal behaviors. In each case 
nearly all the flux is from the fits in the rising limbs. Formally we can 
calculate the correlation coefficient for the rising vs setting rates. As a 
test of this technique we have calculated the correlation coefficient for a two 
day data sample from the Crab. The rise and set data for the Crab are well 
correlated with a correlation coefficient of 0.77 and a corresponding 
probability of $2.6\%$ that the observed correlation could arise from a random 
sample. For the two events shown in Figures~\ref{fig:NGC6752cand} and 
\ref{fig:NGC5139cand} the correlation coefficients are -0.07 and -0.10 
respectively, with a $54\%$ and $49\%$ probability of being random. We have 
calculated the correlation coefficient for the other 11 candidate events 
during the relevant time intervals. For each event the rises and sets either 
show very weak correlation or, in several cases, strong anticorrelation. The 
most significant correlation has a $30.3\%$ probability of being random. The 
lack of significant positive correlations imply that the ``events" observed in 
the light curves do not originate from the target sources but are due to 
interfering sources located elsewhere.

As a check to this interpretation we have analyzed the event in NGC~6752 
shown in Figure~\ref{fig:NGC6752cand} in more depth. We have chosen to look 
further at this particular event because it is somewhat extended in time, 
lasting for approximately 5 days, and reaches moderate flux levels. We have 
produced a map in an approximately $7^{\rm{o}}\times7^{\rm{o}}$ region centered 
on NGC~6752 during the time of the candidate event. The map is created by 
forming a grid of points separated by $0.5^{\rm{o}}$, and producing 
occultation histories for each grid point. The resulting rates for each point 
can be combined together to form a map. The map for the NGC~6752 event shows 
an emission enhancement well localized at $\sim3^{\rm{o}}$ `northeast' of the 
cluster during the first GRO pointing period (TJD~$9078-9083$). It is also 
clear from the limb geometry that such an event would be visible only on the 
setting rates, and indeed this is what is observed in the rate history for 
NGC~6752. During the second pointing period (TJD~$9083-9090$), the emission 
peak moves to $\sim1^{\rm{o}}$ `southwest' of the cluster, and is well 
localized only in the detector with the less optimal pointing. All of this 
leads us to conclude that the enhancement is not associated with the cluster.

\subsection{Recurrence time limits from DXSs}

%absence of hard X-ray emission from clusters
%rule out dim sources as Aql X-1's

\begin{figure*}
\figurenum{3a}
\epsscale{2.1}
\plotone{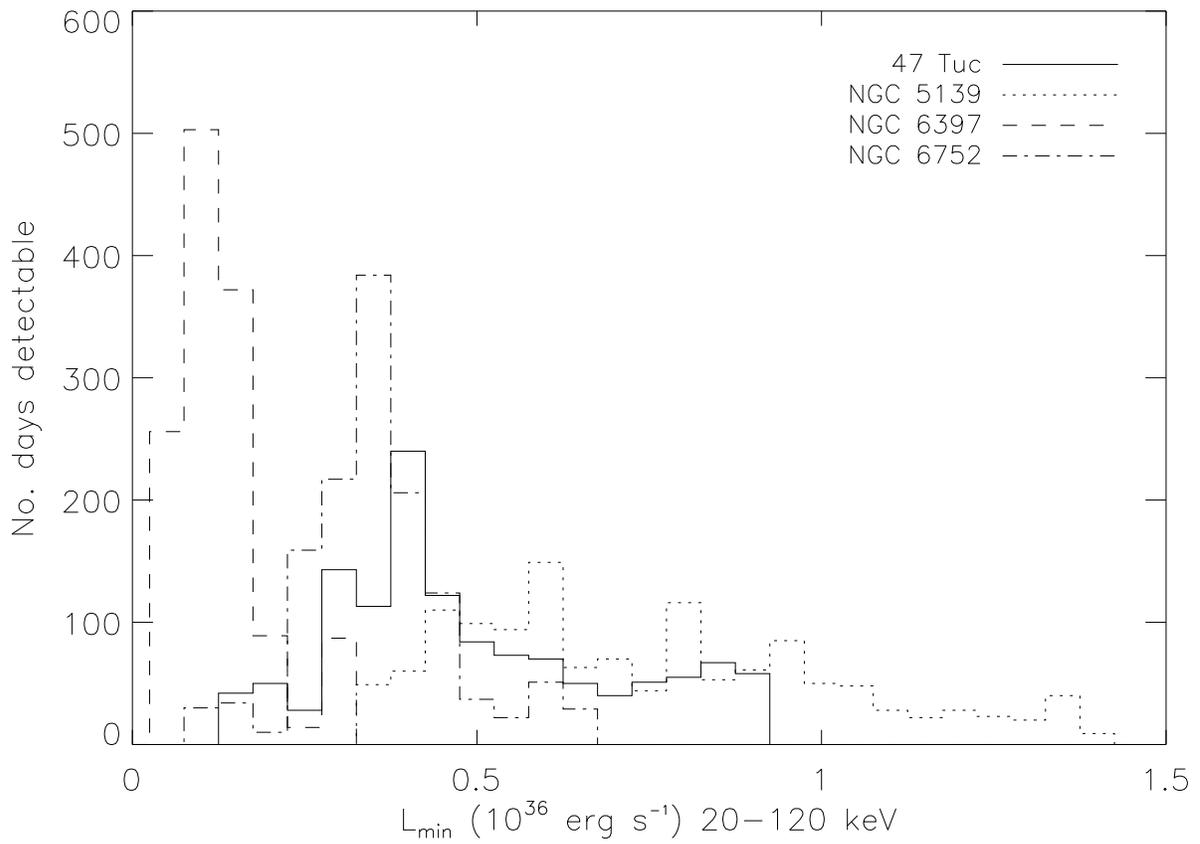}
\caption{Coverage for the four clusters of Search~A. Data integration time 
 is 10.0 days.}
\label{fig:cov_all}
\end{figure*}

\begin{figure*}
\figurenum{3b}
\epsscale{2.1}
\plotone{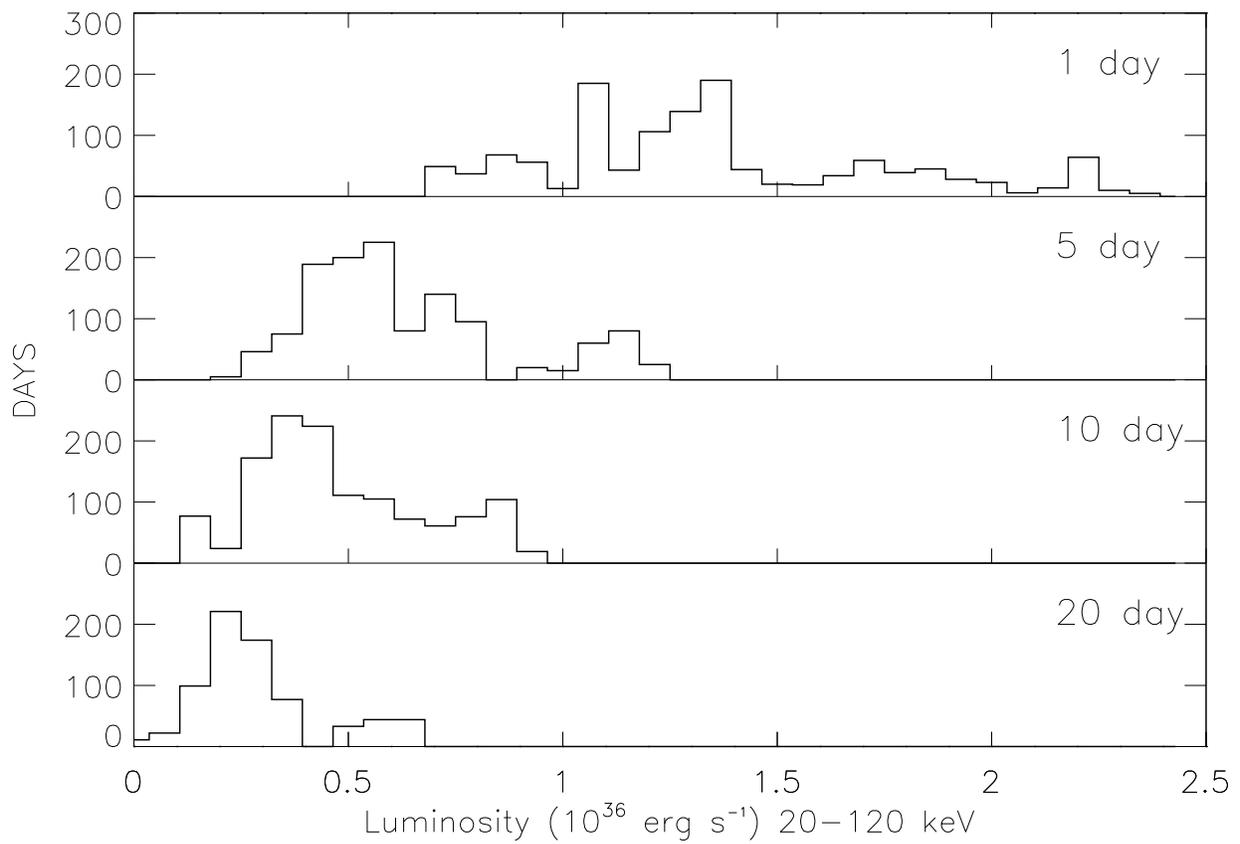}
\caption{Coverage for the cluster 47~Tuc using various data integration times.}
\label{fig:cov_int}
\end{figure*}

In the absence of any detected transient events, we are able to set a lower
limit to the recurrence times of transient sources in the clusters of Search~B. 
For a given time interval in a GC light curve, there will be a minimum 
luminosity, $L_{min}$, at which an event would be visible. $L_{min}$ is 
determined by the width of the flux distribution. We step through the cluster 
light curves and determine the total number of days that events of different 
$L_{min}$s would be observable. The result is a distribution of the number of 
days observable vs $L_{min}$ (Figure~\ref{fig:cov_all}). The $L_{min}$ 
distribution depends on the assumed duration of outburst events and the
photon index used in the deconvolution. This dependence, however, is rather 
weak. The distribution depends more strongly on the time intervals into which 
the data are integrated. Figure~\ref{fig:cov_int} shows the $L_{min}$ 
distribution for different data integration times. The minimum observable 
luminosities are shifted down at longer integration times due to an increased 
sensitivity, i.e. a smaller width in the distribution of fluxes.

\placefigure{fig:cov_all}

\placefigure{fig:cov_int}

\placetable{table:tau_limits}

\begin{table*}[tbp]
\begin{tabular}{|l|l|l|l|l|l|l|}
\tableline
Cluster & $D$ & $N_{DXS}$ \
 & $T_{obs}$ & $\tau_{r}$ & $T_{obs}$ & $\tau_{r}$\\
  & (kpc) & & (years) & (years)  & (years) & (years) \\
 & & & $5\times10^{35}$~erg~s$^{-1}$ & & $1\times10^{36}$~erg~s$^{-1}$ & \
 \\ \tableline
47~Tuc       & 4.6 & 5 & 2.0 &  3.4 &  3.5 &  5.8 \\ \tableline
NGC~5139     & 5.2 & 2 & 0.6 &  0.4 &  2.9 &  1.9  \\ \tableline
NGC~6397     & 2.2 & 5 & 3.6 &  6.0 &  3.6 &  6.0  \\ \tableline
NGC~6752     & 4.1 & 3 & 3.2 &  3.2 &  3.6 &  3.6  \\ \tableline
Total        &     &   & 9.4 & 13.0 & 13.6 & 17.4  \\ \tableline
\end{tabular}
\caption{Recurrence time limits for the GCs in Search I. $T_{obs}$ are the 
integrated times per cluster during which events of the specified luminosity 
would be detectable. $\tau_{r}$ are the corresponding maximum recurrence time 
of outbursts from the dim X-ray sources (DXSs). $N_{DXS}$ are the estimated 
number of known DXSs in each cluster.}
\label{table:tau_limits}
\end{table*}

\begin{figure*}
\figurenum{3c}
\epsscale{2.1}
\plotone{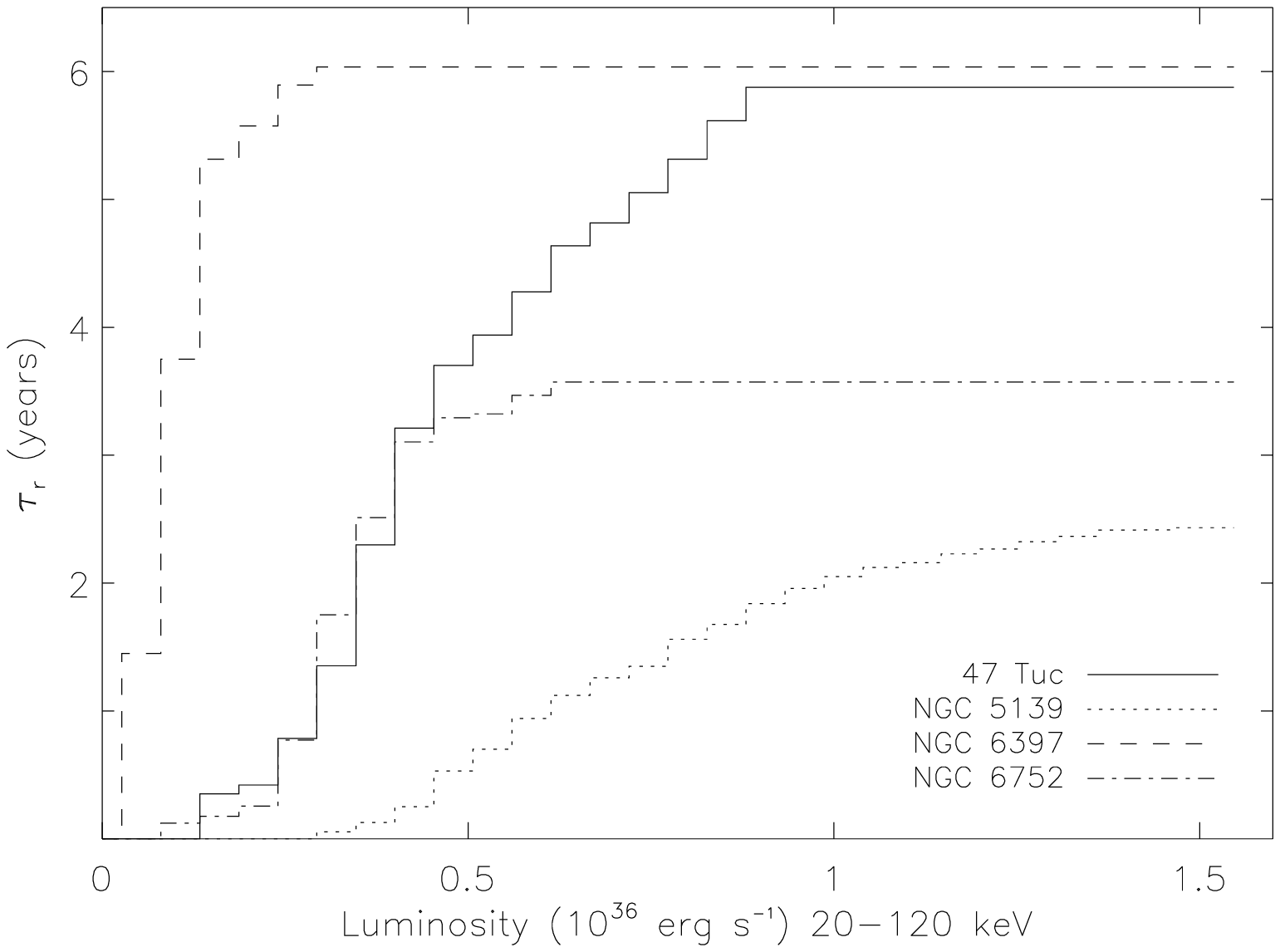}
\caption{Recurrence time lower limits for dim X-ray sources. Recurrence times
are constrained to lie above the curves. A data integration time of 10.0 days 
has been used.}
\label{fig:tau}
\end{figure*}

We determine limits to the recurrence times of outbursts from the $L_{min}$
distribution. For each $L_{min}$ there is an associated observable time, 
$T_{obs}$, for which an event exceeding $L_{min}$ would be detectable. 
$T_{obs}$ is essentially the integral of the $L_{min}$ distribution. $T_{obs}$
is related to the mean event recurrence time, $\tau_{event}$, by 
$P = e^{-T_{obs}/\tau_{event}}$, where $P$ is the probability of observing no
events in time $T_{obs}$ assuming a Poisson distribution of events. 
We calculate a lower limit for $\tau_{event}$ using a 5\% probability of
detection. With multiple X-ray sources in each cluster, an event 
could originate from any of the sources. The minimum event recurrence time, 
$\tau_{event}$, can be stated as a minimum recurrence time for sources, if we 
use the number of DXSs in each cluster, $N_{DXS}$. The minimum source 
recurrence time, $\tau_{r}$, is given by $\tau_{r}=N_{DXS}\cdot \tau_{event}$
(Table~\ref{table:tau_limits}). We also calculate a total recurrence time for 
all the sources as a class using the total $T_{obs}$ for all clusters 
(which gives the total $\tau_{event}$) and the total $N_{DXS}$ of 15. 
Figure~\ref{fig:tau} is plot of $\tau_{r}$ vs $L_{min}$ for the four clusters. 
Recurrence times for sources in the clusters are constrained to lie above the 
lines. It is clear that with increasing event luminosity, $T_{obs}$ increases 
and therefore $\tau_{r}$ increases, giving a tighter recurrence time 
constraint at higher luminosities. Table~\ref{table:tau_limits} displays 
recurrence time lower limits for event luminosities of $5\times10^{35}$ and 
$1\times10^{36}$~erg~s$^{-1}$. The estimates of $N_{DXS}$ are taken from the 
literature: 47~Tuc (Hasinger, Johnston and Verbunt 1994), 
NGC~5139 ($\omega $Cen) (Cool et al. 1995a), 
NGC~6397 (Cool et al. 1993), and NGC~6752 (Grindlay 1993b). For 47~Tuc we take 
only the DXSs within the core. For the latter three clusters we have assumed, 
for consistency, that sources within 5~core radii of each cluster center are 
associated with the GC and others are not. Note that changing $N_{DXS}$ simply 
scales the numbers in Table~\ref{table:tau_limits} and the curves in 
Figure~\ref{fig:tau}.

\placefigure{fig:tau}

These recurrence time limits can be compared to the several recurrent 
transients that have been discovered with all-sky X-ray surveys. One such 
transient is Aql~X-1, observed over a 7 year period with the Vela 5B satellite. 
It has an outburst recurrence time measured at 1.2 years (Priedhorsky and 
Terrell 1984). Another example is Cen~X-4 from which hard X-ray outbursts were 
observed in 1969 and 1979 (Kaluzienski, Holt and Swank 1980, Bouchacourt 
et al. 1984), indicating that its recurrence time is approximately 10 years or 
less, putting it at the outer edge of our limit. Both Aql~X-1 and Cen~X-4 have 
exhibited hard X-ray emission during outburst in excess of 
$1\times 10^{36}$~erg~s$^{-1}$ ($20-120$~keV) (c.f.~Table~\ref{table:hardx}). 
A number of other sources have also displayed recurrent outbursts: 1608-522, 
1630-472, 1730-335 (Van Paradijs and Verbunt 1984, Chen, Shrader and Livio 
1996). These systems have recurrence times of approximately 0.5 to a few years. 

%evidence in favor of CV interpretation

Our recurrence limits indicate that transient sources with outburst 
luminosities greater than approximately $1\times 10^{36}$~erg~s$^{-1}$ 
($20-120$~keV) and recurrence times less than about 2 to 6 years cannot 
constitute the population of dim GC sources in the four sampled clusters. The 
recurrence time lower limit for all sources is $\sim20$~years, if one is 
willing to consider them as a class. This suggests that the dim sources in 
these clusters are not quiescent LMXBs of a type similar to Aql~X-1 or Cen~X-4. 
Some caution should be taken with this interpretation. The outburst recurrence 
times are determined well only for a few systems which may not be typical. 
Also LMXBs in GCs may have systematically different properties than the field 
binaries in which the hard X-ray emission and recurrent outbursts have been 
observed. Indeed some bright LMXBs in GCs have unusual properties, for example 
very short orbital periods (Bailyn 1996). One might argue that a substantial 
number of X-ray outbursts from Aql~X-1-like LMXBs might be detectable only 
below 20~keV. If this were the case, such events would be not detectable by 
BATSE. However, the recently monitored behavior of Aql~X-1 shows clearly that 
BATSE can usually detect hard X-ray emission at times when the optical 
counterpart of Aql~X-1 is excited. These events are most likely associated with 
major X-ray outbursts (Harmon et al. 1996).

\section{Limits from the Persistent Emission Search}

\begin{table*}[tbp]
\begin{tabular}{|l|l|l|l|l|l|}
\tableline
Cluster & $F_{ul}$ & $L_{ul}$ & $N_{isol}$ & %
 $F_{inter}$ & $N_{inter}$ \\
 & ($\gamma$ cm$^{-2}$s$^{-1}$) & (erg s$^{-1}$) & %
      & ($\gamma$ cm$^{-2}$s$^{-1}$) & \\
 & [$\times10^{-4}$] & [$\times10^{34}$] & & [$\times10^{-5}$] & \\ \tableline
47~Tuc       &  3.3 & 5.8 & 19 &  0.71 & 46  \\ \tableline
NGC~5139     &  4.1 & 9.2 & 31 &  0.62 & 66 \\ \tableline
NGC~6397     &  4.2 & 1.7 & 6  &  3.10 & 14 \\ \tableline
NGC~6752     &  3.6 & 5.0 & 17 &  0.85 & 42  \\ \tableline
\end{tabular}
%\end{center}
\caption{Persistent emission upper limits for GCs in Search I. The table lists 
$F_{ul}$ (values are $\sim$1~mCrab), the $2\sigma$ flux upper limits 
($20-120$~keV), $L_{ul}$, the corresponding luminosity upper limits for power 
law spectra with index 2, $N_{isol}$, the maximum number of isolated MSPs 
producing hard X-ray emission based on the model of Chen (1991), $F_{inter}$, 
the estimated flux produced by a single interacting MSP, and $N_{inter}$, the 
maximum number of interacting binary MSPs for an assumed pulsar spin-down 
power of $10^{34}$~erg~s$^{-1}$ (Tavani 1993b).}
\label{table:persist_limits}
\end{table*}

\placetable{table:persist_limits}

We have searched for longer time scale persistent emission using the flux 
histories for the four GCs in Search~B. Previous upper limits have 
been obtained only for 47~Tuc. Observations with SIGMA spanning seven days set 
a limit at $\sim2\times 10^{35}$~erg~s$^{-1}$ ($40-100$~keV) 
(Barret et al. 1993) and an observation by the balloon-borne EXITE instrument 
found an upper limit of $1.4\times 10^{36}$~erg~s$^{-1}$ ($30-150$~keV) 
(Grindlay et al. 1993c). Upper limits have also been obtained for 47~Tuc with 
COMPTEL, OSSE and EGRET in their respective bands (O'Flaherty et al. 1995). 

Our BATSE search offers improved sensitivity as a result of the very long
integration time. We have detected no emission for the four clusters. The 
upper limits are summarized in Table~\ref{table:persist_limits}. We have 
obtained these upper limits by formal error calculations integrating over
large time windows. As discussed in Section 3, such a procedure yields error 
values systematically larger than the actual flux distribution. We have studied 
this effect for the four GCs of Search~B and determined a correction factor
of 1.6 based on fitted widths of flux distributions. We have used the factor of 
1.6 correction to the formal error estimates for the upper limit values of 
Table~\ref{table:persist_limits}. The clusters of Search~B were chosen to limit 
the systematic effects of interfering sources, a selection which permits very 
sensitive flux limits. We note that the sensitivities may depend on the assumed 
spectral shape used in deconvolution. We have employed power law spectra with 
photon index 2.0.

%\placetable{table:persist_limits}

Using our upper limits, we constrain the number of isolated high energy 
emitting MSPs in the clusters. The luminosity for a typical non-interacting 
MSP can be calculated. Chen (1991) does this using the outer-gap emission 
model of Cheng, Ho and Ruderman (1986) and the observed period distribution 
for MSPs. The resulting luminosity from a cluster is a constant times the 
number of isolated MSPs, assuming isotropic emission. For a cluster with one 
MSP, this would yield a luminosity of approximately 
$3\times 10^{33}$~erg~s$^{-1}$. Combining this luminosity estimate with our 
luminosity upper limit, we find the maximum number of isolated MSPs, 
$N_{isol}$, in each of the clusters (Table~\ref{table:persist_limits}). 
The estimate of $N_{isol}$ depends on the period distribution of MSPs.  In 
the luminosity calculation, Chen (1991), uses the period distribution of all 
the 28 then-known MSPs, fit as a falling power law. This distribution is 
poorly constrained due to selection effects and small number statistics and 
may vary from cluster to cluster. This adds additional uncertainty to our 
values of $N_{isol}$.

The flux from an interacting binary MSP has been estimated by Tavani (1993b).
Using this model we calculate the estimated flux from each cluster for a 
single interacting MSP system, $F_{inter}$ (Table~\ref{table:persist_limits}). 
Combined with our flux upper limits, the flux estimate yields an upper limit on 
the number of interacting MSPs, $N_{inter}$. The constraints on $N_{inter}$ 
are weak. This calculation uses an average MSP spin-down power of 
$10^{34}$~erg~s$^{-1}$ (Taylor et al. 1995). MSP spin-down powers are in the 
range $10^{33}$-$10^{36}$~erg~s$^{-1}$ but are still not well constrained.

\section{Conclusions}

We have monitored 27 galactic globular clusters (GCs) with BATSE during a 
period of approximately four years each. We have detected no distinct hard 
X-ray outburst episodes and no persistent emission. The lack of detected 
events has several interesting implications.

For the GCs with dim X-ray sources (DXSs), we find a lower limit for the 
outburst recurrence time from DXSs of $\tau_{r}\sim2-6$~years.
This limit excludes the existence of `Aql~X-1-like' objects (i.e., 
persistent X-ray sources subject to major X-ray outbursts with a time scale 
of $\sim 1$ year), since $\tau_{r}$ is comparable to or greater than this
outburst recurrence time scale. This suggests that the DXSs in these clusters
are not LMXBs similar to Aql~X-1.

The lack of strong outburst events in our search means we also have no 
evidence of accreting BHs in GCs. If active BH binaries exist in GCs, they 
would be clearly detectable with hard X-ray luminosities of order of 
$10^{36}-10^{37}$~erg~s$^{-1}$.

We have calculated upper limits on persistent hard X-ray emission from GCs.
The limiting luminosity is $\sim(2-10)\times10^{34}$~erg~s$^{-1}$ for the 
closest clusters. The limit for 47~Tuc is somewhat more sensitive than previous 
measurements. A model-dependent but reasonable estimate of the expected hard 
X-ray magnetospheric emission from isolated MSPs (Chen 1991) implies that the 
number of isolated MSPs emitting hard X-rays in 47~Tuc is less than 19. Taking 
the observed 11 MSPs in 47~Tuc, and a beaming factor of 2 for short period 
pulsars, this upper limit is comparable to actual number of MSPs in 47~Tuc. 
Our results are therefore a constrain the magnetospheric model of MSP hard 
X-ray emission.

Our upper limits on persistent hard X-ray emission also provide a mild
constraint on the number of interacting pulsars in binaries. The observed 
efficiency (a few percent) of conversion of pulsar spindown luminosity into 
hard X-ray emission in the case of the periastron passage of the Be star/pulsar 
system PSR~B1259-63 (Tavani et al. 1996, Grove et al. 1995) together with the 
observed MSP spindown average luminosity of $10^{34}$~erg~s$^{-1}$ implies 
a limit to the GC population interacting pulsar binaries. We find limits of 
$14-66$ interacting MSPs in the clusters studied.

\acknowledgments

We would like to acknowledge the BATSE instrument team for their support.
We thank Jonathan Grindlay, Didier Barret and Andrew Chen for helpful comments. 
We thank Mark Stollberg for assistance with data analysis. This work is
Columbia Astrophysics Laboratory Contribution Number 596. This work is
supported in by NASA Grants NAG~5-2235 and NGT~8-52806.

%FIGURES%%%%%%%%%%%%%%%%%%%%%
%%%% the following are the same figures scaled for the preprint size (aas2pp4)

\clearpage

%
%%%%%%%%%%%%%%%%%%%%%%%%%%%


\begin{references}

\reference{bahcall75} Bahcall, J. and Ostriker, J. 1975, Nature, 256, 23
\reference{bailyn96} Bailyn, C. 1996, in The Origins, Evolution, and 
  Destinies of Binary Stars in Clusters, eds. E.F. Milone and J.C. Mermilliod,
  (ASP Conf. Ser.), in press
\reference{bailyn95} Bailyn, C. 1995, ARAA, 33, 133
\reference{barret95} Barret, D. and Grindlay, J. 1995, \apj, 440, 841
\reference{barret96} Barret, D. et al. 1996, in 3rd Compton Symposium, 
 Munich, in press
\reference{barret94} Barret, D. and Vedrenne, G. 1994, ApJS, 92, 505
\reference{barret93} Barret, D. et al. 1993, \apj, 405, L59
\reference{barret92} Barret, D. et al. 1992, \apj, 394, 615
\reference{barret91} Barret, D. et al. 1991, \apj, 379, L21
\reference{bouchacourt84} Bouchacourt, P. et al. 1984, \apj, 285, L67
\reference{chen91} Chen, K. 1991, Nature, 352, 695
\reference{chen93} Chen, K., Middleditch, J. and Ruderman, M. 1993, 
  \apj, 408, L17
\reference{chen96} Chen, W., Shrader, C., and Livio, M. 1996, \apj, submitted
\reference{cheng86} Cheng, K.S., Ho, C., and Ruderman, M. 1986, \apj, 300, 500
\reference{churazov95} Churazov, E., et al. 1995, \apj, 443, 341
\reference{churazov94} Churazov, E., et al. 1994, Adv. in Space Res., COSPAR,
  Munich, in press
\reference{clark75} Clark, G.W. 1975, \apj, 199, L143
\reference{cool95a} Cool, A.M. et al. 1995a, \apj, 438, 719
\reference{cool95b} Cool, A.M. et al. 1995b, \apj, 439, 695
\reference{cool93} Cool, A.M. et al. 1993, \apj, 410, L103
\reference{forman76} Forman, W., Jones, C. and Tananbaum, H. 1976, \apj, 
 207, L25
\reference{grindlay93a} Grindlay, J.E. 1993a, Adv. Space Res., 13(12), 597
\reference{grindlay93b} Grindlay, J. 1993b, in Dynamics of Globular Clusters - 
 eds. S. Djorgovski and G. Meylan (ASP Conf. Ser.), 285
\reference{grindlay93c} Grindlay, J. et al. 1993c, in Compton Gamma-Ray 
 Observatory, AIP Conf. Proc. 280, eds. M.Friedlander, N.Gehrels and D. Macomb,
 (New York:AIP), 243
\reference{grove95} Grove, J.E., et al. 1995, \apj, 447, L113
\reference{harmon96} Harmon, B.A. et al. 1996, in 3rd Compton Symposium, 
 Munich, in press
\reference{harmon92} Harmon, B.A., et al. 1992, Compton Observatory Science
 Workshop, NASA CP 3137, 69
\reference{hasinger94} Hasinger, G., Johnston, H.M., and Verbunt, F. 1994, 
 \aap, 288, 466
\reference{HG83} Hertz, P. and Grindlay, J. 1983, \apj, 275, 105
\reference{hertz85} Hertz, P. and Wood, K.S. 1985, \apj, 290, 171
\reference{hut92} Hut, P. et al. 1992, PASP, 104, 981
\reference{kulkarni93} Kulkarni, S.R., Hut, P. and McMillan, S. 1993, 
  Nature, 364, 421
\reference{kaluzienski80} Kaluzienski, L.J., Holt, S.S., and Swank, J.H.
  1980, \apj, 241, 779
\reference{kluzniak88} Kluzniak, W. et al. 1988, Nature, 334, 225
\reference{lyne96} A. Lyne, 1996, in Proc. 7th M. Grossmann Symposium, eds. M. 
 Kaiser, R. Jantzen, (World Scientific), in press
\reference{margon87} Margon, B. and Bolte, M. 1987, \apj, 321, L61
\reference{mereghetti95} Mereghetti, S. et al. 1995, \aap, 302, 713
\reference{mitsuda89} Mitsuda, K. et al. 1989, PASJ, 41, 97
\reference{oflaherty95} O'Flaherty, K.S., et al. 1995, \aap, 297, L29
\reference{pendleton96} Pendleton, G.N. et al. 1996, NIM, in press
\reference{peterson93} Peterson, C.J. 1993, in Structure and Dynamics of 
 Globular Clusters: ASP Conf. Proc. Vol. 50, eds. S.G.Djorgovski and G. Meylan
 (SanFrancisco:ASP), 337
\reference{priedhorsky94} Priedhorsky, W.C. and Terrell 1984, \apj, 280, 661
\reference{pye83} Pye, J. and McHrady, I. 1983, MNRAS, 205, 875
\reference{robinson95} Robinson, C. et al. 1995, MNRAS, 274, 547
\reference{ruderman88} Ruderman, M. and Cheng, K.S. 1988, \apj, 335, 306
\reference{tanaka95} Tanaka, Y. and Lewin, W.H.G. 1995 in X-Ray Binaries,
  eds. W.H.G.Lewin, J.van Paradijs, and E.P.J. 
  van den Heuvel (Cambridge University Press), 126
\reference{tanaka89} Tanaka, Y. 1989, in Proc. of the 23rd ESLAB Symposium,
  Vol. 1, ed. J.Hunt (ESA Pub. Div.), 3
\reference{tavani96} Tavani M., et al., 1996, A\&AS, in press
\reference{tavani93a} Tavani, M. 1993a, AAS, 97, 313
\reference{tavani93b} Tavani, M. 1993b, \apj, 407, 135
\reference{tavani91} Tavani, M. 1991, \apj, 379, L69
\reference{tomaney94} Tomaney, A., Crotts, A., and Shafter, A. 1994, 
 BAAS, 181, 73.09
\reference{taylor95} Taylor, J. et al. 1995, Princeton Pulsar Catalog
\reference{vanParadijs84} Van Paradijs, J. and Verbunt, F. 1984,
 in High Energy Transients in Astrophysics, ed. S.E. Woosley (New York: AIP), 49
\reference{verbunt84} Verbunt, F., van Paradijs, J., and Elson, R. 1984, 
  \mnras, 210, 899
\reference{verbunt94} Verbunt, F., et al. 1994, MmSAI, 65, 249
\reference{verbunt95} Verbunt, F., et al. 1995, \aap, 300, 732
\reference{white84} White, N.E., Kaluzienski, J.L., and Swank, J.H. 1984, 
 in High Energy Transients in Astrophysics, ed. S.E. Woosley (New York: AIP), 31
\reference{zhang96} Zhang, S.N. 1996, in 3rd Compton Symposium, 
 Munich, in press
\reference{zhang93} Zhang, S.N., et al. 1993, Nature, 366, 245

\end{references}
\end{document}